**Knowing You Is Everything: LLM Agents Achieve Near-Perfect Profile-Consistent Reaction Prediction in Social Media Simulation**

Ljubiša Bojić[1,2,3,*], Ph. D., Senior Research Associate
(Corresponding author; Email address: ljubisa.bojic@ivi.ac.rs; ORCID: 0000-0002-5371-7975)

Ljiljana Matić[4], Ph. D. Candidate, Teaching Assistant
(Email address: ljiljana.matic@ef.kg.ac.rs; ORCID: 0009-0002-0473-0214

Jörg Matthes[5], Ph. D., Full Professor,
(Email address: joerg.matthes@univie.ac.at; ORCID: 0000-0002-5371-7975)

Milan Čabarkapa[6], Ph. D., Assistant Professor
(Email address: mcabarkapa@kg.ac.rs; ORCID: 0000-0002-2094-9649)

Bojana Dinić[7], Ph. D., Full Professor
(Email address: bojana.dinic@ff.uns.ac.rs; ORCID: 0000-0002-5492-2188)

Jue Wang[8], Ph. D., Associate Professor
(Email address: wangjue@ntu.edu.sg; ORCID: 0000-0002-3401-713X)

[1]Institute for Artificial Intelligence Research and Development of Serbia; Address of correspondence: 1 Frukogorska 1, 21000 Novi Sad, Serbia;
[2]University of Belgrade, Institute for Philosophy and Social Theory, Digital Society Lab; Address of correspondence: Kraljice Natalije 45, 11000 Belgrade, Serbia;
[3]Complexity Science Hub, Vienna, Austria; Address of correspondence: Metternichgasse 8, 1030 Vienna, Austria;
[4] University of Kragujevac, Faculty of Economics, Kragujevac, Serbia; Address of correspondence: Sestre Janjić 6, 34000 Kragujevac, Serbia;
[5] University of Vienna, Faculty of Social Sciences, Department of Communication; Address of correspondence: Währinger Straße 29 (R. 7.47), 1090 Vienna, Austria;
[6] University of Kragujevac, Faculty of Engineering, Kragujevac, Serbia; Address of correspondence: Sestre Janjić 6, 34000 Kragujevac, Serbia;
[7]University of Novi Sad, Faculty of Philosophy, Novi Sad, Serbia; Address of correspondence: Dr Zorana Đinđića 2, 21102 Novi Sad, Serbia;
[8] Nanyang Technological University, School of Social Sciences, Singapore; Address of correspondence: 50 Nanyang Ave, 639798 Singapore;

## Abstract

Autonomous AI agents in social media present concrete risks to democratic discourse and platform governance, while also offering tools for pre-deployment recommender system testing. A central open question is whether persona-prompted LLMs can simulate individual-level social media reactions with sufficient accuracy to support either application, and how accuracy depends on profile completeness, model selection, and the generalization challenge posed by novel post content. This study benchmarks twelve LLM configurations on binary like/dislike prediction across 296 survey-based agent profiles and 26 ground-truth-mapped posts under three profile conditions, with leave-post-out machine learning classifiers as baselines. Across full-profile conditions, accuracy ranges from 75.54% to 96.68%, with a 30-point spread attributable primarily to model selection and confirmed by paired McNemar tests with agent-level bootstrap intervals. GPT-5.5 Pro accuracy degrades monotonically from 96.68% under a full profile to 62.32% under a reduced profile and to 51.00% with demographics alone, the last indistinguishable from the majority-class baseline, which confirms that demographic inference provides negligible predictive signal. Supervised classifiers collapse to 15.4% under leave-post-out, while LLMs sustain genuine zero-shot generalization unavailable to trained methods. Adaptive reasoning improves accuracy substantially for some models. Inter-model agreement is nearly double for posts with direct profile anchors (mean κ = 0.44) than for posts without them (κ = 0.23), and the least heterogeneous configuration homogenizes 34% of simulated population reactions. Results validate LLM-based simulation for recommender system stress-testing while documenting the behavioral accuracy that makes large-scale synthetic agent swarms a credible threat to public opinion.

## Introduction

Social media platforms have become the primary infrastructure through which billions of people encounter political information, form opinions, and signal their preferences to others (Bojic, 2024). The behavioral patterns that emerge from this infrastructure, which posts receive engagement, which framings provoke dislike, and which kinds of content spread across social networks, are no longer governed solely by human choices. Algorithmic systems mediate what content reaches which users, and an expanding class of AI-generated actors participates directly in online discourse, producing posts, comments, and reactions indistinguishable to casual observers from those of human users (Schroeder et al., 2026). The societal stakes of this development are difficult to overstate. Recommendation systems have been shown to shift political attitudes at scale. Recent evidence from X (formerly Twitter) demonstrates that algorithmic feed design can produce systematic shifts in political orientation among large user populations (Gauthier et al., 2026). Swarms of coordinated AI agents can manufacture the appearance of grassroots consensus, adapting their engagement strategies in real time to maximize persuasive impact (Schroeder et al., 2026). At the same time, these same technologies offer genuine opportunities for platform governance research, allowing investigators to test the downstream consequences of recommendation algorithm changes in simulation before exposing real users to them (Grossmann et al., 2023). Whether AI agents are deployed for harm or for constructive governance research, the quality of their behavioral output depends on the same underlying property: how faithfully they can replicate the reactions of the specific kinds of human beings they are meant to represent. That question of individual behavioral fidelity is the central concern of the present study.

The idea that large language models (LLMs) can serve as proxies for human participants has moved rapidly from theoretical proposal to research practice. Park et al. (2023) provided an early demonstration that LLM-powered generative agents could produce socially credible behavior in a sandbox environment, with human observers judging their interactions as believable. That proof of concept opened a research program that has since grown substantially in ambition and scale. Park et al. (2026) grounded agents in structured interviews with over a thousand real individuals and found that the resulting agents could reproduce those individuals' responses on a broad social survey at a rate approaching the consistency with which the participants themselves agreed with their own earlier answers when tested two weeks later. Yang et al. (2024) extended the simulation approach to one million agents, producing emergent dynamics of information spreading and group polarization in environments modeled on Reddit and X. In a complementary direction, Altera.AL (2024) showed that LLM-powered agents could autonomously develop social roles, democratic governance structures, and even cultural transmission across multi-agent environments, suggesting that emergent social organization is achievable without explicit programming of societal architecture. Guo et al. (2024) found that that multi-agent LLM systems can be deliberately structured to behave coherently at the collective level. These results establish LLM-based simulation as a serious methodological option for the social sciences, capable in principle of testing interventions and modeling dynamics that would be impractical or ethically problematic to study with real participants (Grossmann et al., 2023).

The dominant method for conditioning an agent's behavior toward a specific individual is persona prompting, in which a textual description of the target person's demographic, attitudinal, and psychological profile is supplied to the model as part of its input. The approach was given early validation by Argyle et al. (2023), who showed that conditioning GPT-3 on demographic profiles drawn from a large-scale US electoral study produced opinion distributions that tracked actual subgroup responses in that study, a pattern they characterized as "silicon sampling." Törnberg et al. (2023) applied persona-based agents to a simulated social media platform and used the resulting system to compare different news feed algorithms, finding that a bridging algorithm produced more cross-partisan dialogue than conventional engagement-maximizing designs.

Subsequent research has complicated this picture. Hu & Collier (2024) measured how much variance in human annotation behavior could be attributed to persona variables across a range of NLP tasks and found that the contribution was consistently small, below ten percent in most task conditions, with gains concentrated in settings where human disagreement was already moderate. Liu et al. (2024a) found that agents steered toward attitudinally incongruent personas, for example a political liberal who holds positions typically associated with the right, were nearly ten percentage points less accurate than agents given congruent personas, and that models tended to default to the stereotypical stance for a demographic category rather than the specific stance specified in the profile. Cheng et al. (2023) showed that LLM simulations of social identities tend to produce caricatured rather than authentic representations, collapsing the internal variation of a demographic group into a single modal type. La Cava and Tagarelli (2025) found that open-source LLMs varied substantially in their capacity to maintain persona consistency across different task contexts, calling into question whether persona conditioning produces a stable behavioral agent or merely a local modulation of output style. These findings suggest that the relationship between persona description and agent behavior is more fragile than the early silicon sampling results implied, and that models may be performing demographic stereotyping rather than genuine integration of the specific attitudinal information supplied in a profile.

A more fundamental challenge concerns whether the behavioral outputs of LLMs reflect genuine reasoning about a described person or statistical regularities absorbed from training data. Dillion et al. (2023) argued that LLMs may reproduce correct behavioral responses through memorization of training-set patterns rather than through any capacity for individual-level behavioral inference. Qi et al. (2024) provided supporting evidence that LLMs can produce correct outputs accompanied by incorrect explanations, a pattern consistent with retrieval rather than comprehension. Binz et al. (2025), working from a different angle, demonstrated that a foundation model trained at scale on human behavioral data could predict cognitive performance across a diverse range of experimental paradigms at human-level accuracy, which suggests that LLMs encode behavioral regularities at a level of depth that exceeds simple pattern recall, without resolving whether this depth extends to individual-level prediction from a persona prompt. Cloud et al. (2026) showed that behavioral traits can transfer between LLMs through what they describe as subliminal learning, with student models acquiring traits from training data produced by a differently-disposed teacher model, even when the transmission content was semantically unrelated to the target trait. That finding implies that LLMs carry behavioral dispositions that are not fully captured or controlled by explicit persona instructions, which has direct consequences for how simulation researchers should interpret the outputs of persona-prompted agents.

The accuracy of LLM-based simulation varies substantially across content domains, and understanding where that variation falls is essential for any application that depends on behavioral fidelity across diverse post types. Bojić et al. (2025a) showed that GPT-4 can outperform human participants in tasks requiring contextual interpretation of language, suggesting strong capability in domains that reward inferential processing of text. Prior work from the same research group demonstrated that agent-based social media simulations using LLMs can reproduce polarization dynamics associated with recommender system personalization, with higher levels of algorithmic filtering amplifying both affective and structural polarization in simulated networks (Bojić et al., 2025b). Alipour et al. (2024) showed that injecting AI-generated actors with pro-social instructions into a simulated social network could reduce echo chamber formation, which illustrates the potential for simulation-based governance research.

The question of profile completeness has received relatively little systematic attention in the simulation literature. Most studies compare agents with persona information against agents without any persona, but few vary the depth and specificity of profile information across multiple conditions to determine how prediction degrades as the profile becomes thinner.

Münker et al. (2026) argued that the empirical realism of generative agents in social network simulations cannot be assumed from model capability rankings and must be established through direct behavioral benchmarking. Schwager et al. (2026) evaluated conditioned comment prediction across several LLM configurations and found that supervised fine-tuning approaches substantially outperformed zero-shot persona prompting, raising the question of how much of the performance ceiling can be reached without task-specific training.

An underexplored dimension of LLM social simulation is what happens to model predictions across post types for which no ground-truth behavioral data exist. In any realistic simulation scenario, agents will be exposed to content that falls outside the domain of validated survey items. Whether models behave consistently with each other in such cases, and whether they produce heterogeneous reactions that reflect the diversity of the simulated population, are empirical questions with direct implications for the validity of downstream simulation outputs. Serapio-García et al. (2025) found that LLMs can replicate Big Five personality trait profiles but that the correlations among personality dimensions in LLM outputs are unrealistically high relative to human data, suggesting that models represent personality at a coarser level than human populations exhibit. Bodroža et al. (2024) showed that LLMs tend to display socially desirable personality profiles with elevated agreeableness and reduced Machiavellianism, which implies a systematic pull toward consensus rather than genuine population-level heterogeneity. These findings raise concern about the ecological validity of simulations in which agents are assumed to exhibit realistic behavioral variation across all content domains. If models collapse to uniform responses for content outside their anchoring domain, the simulated population will lack the distributional properties of the real one.

The dual-use nature of behavioral simulation frames this work. LLMs have demonstrated persuasive influence. Costello et al. (2024) showed a twenty-percent reduction in conspiracy belief after brief dialogue, with effects lasting two months. Salvi et al. (2025) found similar results in naturalistic conversations and Steyvers et al. (2025) showed that people overestimate LLM calibration. Weidinger et al. (2021) highlighted the ethical risks of deploying such systems. Any capacity to predict individual reactions can also enable targeted content design, raising concerns reflected in the EU AI Act's ban on manipulative AI (Bentzen, 2025) and the DSA's systemic risk rules. Mapping the limits of LLM predictive accuracy is therefore a governance as well as methodological contribution.

Based on the above presented research findings this study poses the following hypotheses and research questions. We hypothesize that prediction accuracy will be substantially above chance for the full profile condition and that accuracy will degrade monotonically as profile information is reduced from full to partial to demographics only, with the demographics-only condition approaching the chance baseline (H1). We hypothesize that positively-framed posts will be predicted more accurately than negatively-framed posts across model configurations (H2), and that posts anchored in survey items with binary preference will reach higher accuracy than posts anchored in Likert-type scales (H3).

We pose as research questions whether adaptive reasoning modes consistently improve accuracy over standard modes within the same model family (RQ1), how substantially model selection affects accuracy across the new generation of frontier models tested here (RQ2), and whether inter-model agreement on the 30 posts without ground-truth survey mappings is systematically lower than agreement on the 26 posts with ground-truth mappings, such that posts in contested political and social domains produce the lowest convergence across models (RQ3). Finally, we examine whether the reaction heterogeneity produced by different models, measured as the distributional spread of like rates across the 296-agent population, varies systematically across model configurations and post types, and whether any configurations produce levels of behavioral homogeneity that would call into question their suitability for population-level social simulation (RQ4).

## Methodology

This study combines a survey-based behavioral dataset collected from 296 Serbian residents with a systematic benchmarking of twelve model configurations on binary like/dislike reaction prediction. A schematic overview of the full procedure is presented in Figure 1.

### *Participants and Procedure*

Survey data were collected from Serbian residents through Latenta, a market research agency, using Alchemer as the survey aggregation platform. Data collection took place between 20 and 23 February 2026. A total of 1012 survey attempts were recorded, of which 323 were fully completed, 311 were partially completed, and 378 participants were screened out based on predefined eligibility criteria. After data cleaning, which removed 643 responses due to incomplete data or identifiable quality issues, the final analytic sample comprised 296 participants (52.4% female, 47.0% male, 0.7% other), aged 18 to 72 years (M = 34.71, SD = 14.17). The sample was distributed across four Serbian regions: Belgrade (n = 80, 27.0%), Vojvodina (n = 80, 27.0%), Šumadija and Western Serbia (n = 79, 26.7%), and Southern and Eastern Serbia (n = 57, 19.3%).

Educational attainment in the sample was diverse. The largest group had completed secondary school (n = 95, 32.1%), followed by participants currently enrolled in bachelor-level studies (n = 42, 14.2%), those who had completed a higher vocational school degree (n = 42, 14.2%), and bachelor graduates (n = 40, 13.5%). Participants currently attending secondary school accounted for 7.8% of the sample (n = 23), while 7.4% held a master's degree (n = 22). Smaller proportions were currently enrolled in higher vocational programmes (3.4%), currently completing master's studies (3.0%), had completed only primary school (0.7%), or had not completed primary school (0.7%). Doctoral degree holders represented 1.4% of the sample (n = 4), and 1.7% reported a different educational background. Participants were recruited with the intent of obtaining a geographically stratified sample representative of the Serbian adult population across age, gender, region, and education categories. Participants accessed the survey online and completed it in Serbian.

*Survey Measures*

Survey items covered five broad domains. The first domain was basic sociodemographic information: age, gender, educational attainment, employment status, and region of residence. The second domain covered institutional trust, with items asking participants to rate their level of trust in the following institutions using a five-point Likert scale ranging from *I do not trust them at all* to *I completely trust them*. The third domain covered attitudes about geopolitical positions on the Russia-Ukraine conflict, EU integration, and the two EU-related belief statements included in the post set with the five-point Likert scale ranging from *I completely disagree* to *I completely agree*. The fourth domain covered entertainment and lifestyle preferences using multiple-response responses indicating preferences for specific movie and music genres and leisure-time activities. The fifth domain covered items referred to personality traits, political orientation and similar individual differences.

*Social Media Posts*

The first category (n = 28) covered news and politics, including posts addressing institutional trust, geopolitical positions, attitudes toward European integration, domestic political developments, and related topics. The second category (n = 28) covered entertainment and lifestyle, addressing everyday preferences and activities including media consumption, genre preferences, music tastes, and leisure habits. Posts were also classified by the valence of their framing. Positively-framed posts presented a topic or preference in an affirming or supportive manner, and negatively-framed posts presented the same or an equivalent topic in a critical, dismissive, or skeptical manner. Among the 26 evaluated posts, 12 were positively framed and 14 were negatively framed.

Within the 56 posts, 26 had a direct correspondence to a survey item from 2-4 domains in the participant's profile, hereafter referred to as evaluated posts. These posts covered binary genre and preference items (science fiction, mystery, books, video games, classical music, pop music), trust-scale items (media, science, state institutions, police, Serbian Orthodox Church), and the agreement-scale geopolitical and EU attitude items (Russia's military justification, Ukraine's peace negotiation conditions, EU market access benefits, EU membership conditions as a threat to national identity). The remaining 30 posts, hereafter referred to as unevaluated posts, covered topics including student protests, the Israel-Palestine conflict, attitudes toward specific political figures, China, trade unions, non-governmental organizations, ecological awareness, and everyday lifestyle activities such as cooking, shopping, volunteering, and interior design. These posts had no corresponding survey items and thus no ground-truth labels for individual-level accuracy evaluation. Across the 26 evaluated posts, 12 were positively framed and 14 were negatively framed. Among the negatively-framed posts, 11 used the same underlying profile column as their positively-framed counterpart (for example, ERN6 and ERP6 both map to the binary classical music preference item), and 1 had a unique corresponding column encoding a negative stance directly (NRN6, mapping to the EU membership threat statement). This distinction is consequential for the evaluation rule system, which is described in the next section.

Posts were written in colloquial Serbian appropriate for social media contexts. All posts were reviewed for clarity, authenticity of register, and balance of representation across political and social perspectives.

Figure 1. Schematic overview of the study design and analysis pipeline.

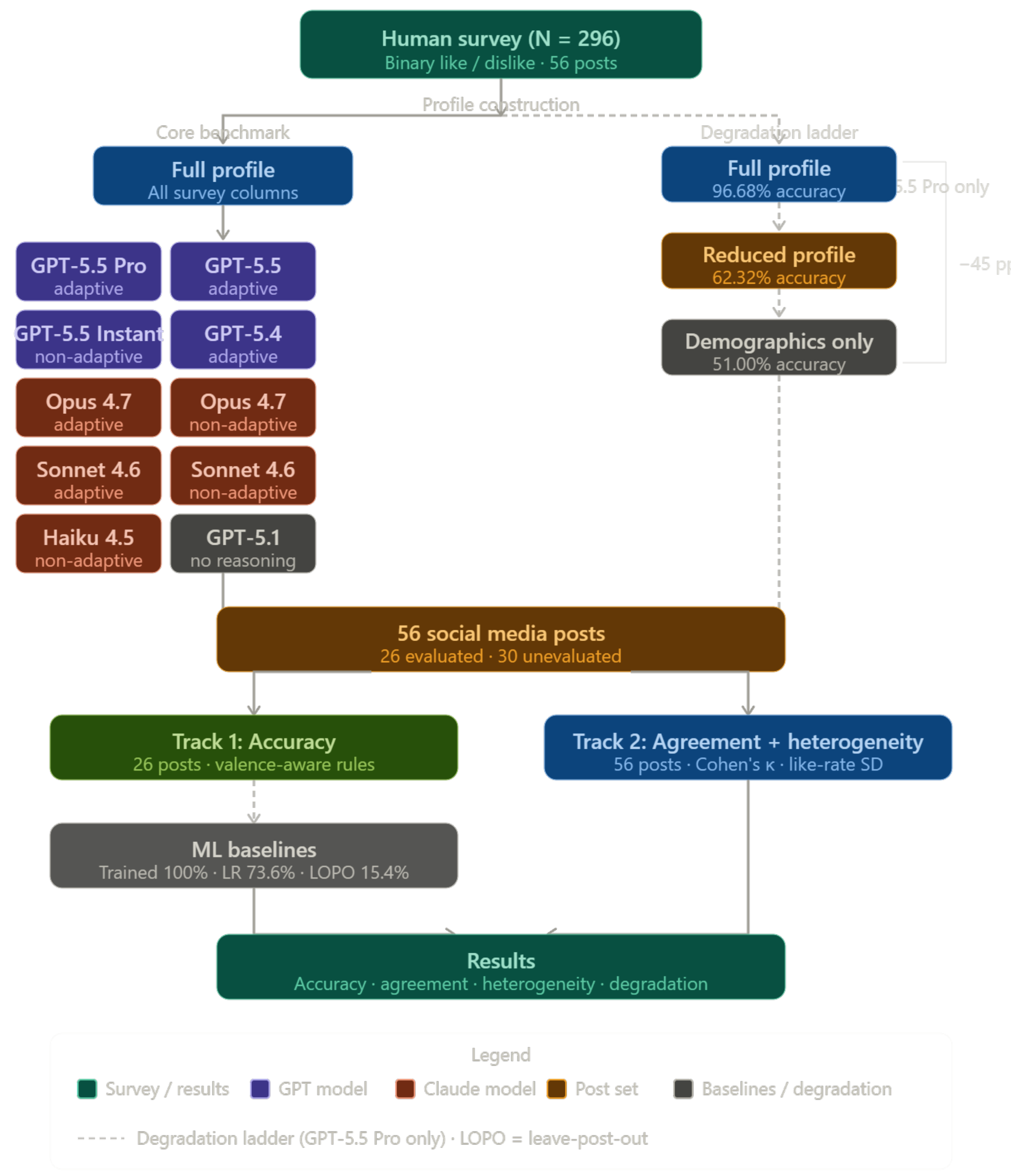

*Agent Construction and Persona Formulation*

For each of the 296 participants, a structured natural-language persona description was constructed by translating survey responses into a formatted English-language profile. Survey responses were translated into English to standardize the input language across all LLM configurations and avoid confounding model performance with cross-lingual processing differences. Three profile conditions were constructed. The full profile condition included all survey information relevant to the post set: demographics, political orientation, institutional trust ratings, geopolitical opinion items, and binary entertainment and media preferences. The reduced profile condition omitted the attitudinal and opinion items most directly related to the evaluated posts (the agreement-scale and trust-scale items), retaining demographic characteristics, personality-adjacent items, and broader behavioral preferences. The demographics-only condition included only age, gender, region of residence, and educational attainment. These three conditions were crossed with model selection to produce the full set of experimental configurations. The full profile condition was applied to all ten core model configurations. The and reduced and demographics-only conditions were applied to the single best-performing model, GPT-5.5 Pro, to establish the lower bound of prediction accuracy achievable from demographic information alone.

*LLM Configurations and Inference Setup*

Twelve LLM configurations constituted the core comparison set. These were drawn from two model families: OpenAI (GPT-5.5 Pro adaptive, GPT-5.5 adaptive, GPT-5.5 Instant non-adaptive, GPT-5.4 adaptive and Anthropic (Claude Opus 4.7 adaptive, Claude Opus 4.7 non-adaptive, Claude Sonnet 4.6 adaptive, Claude Sonnet 4.6 non-adaptive, Claude Haiku 4.5 non-adaptive). Within each family, configurations varied along two dimensions: model version or tier and inference mode. The distinction between adaptive and non-adaptive inference modes corresponds to whether the model operated with extended reasoning or chain-of-thought capabilities activated.

All LLM models were queried via API under identical prompt structures. The data collection took place between 18 and 22 of May 2026 (OSF, 2026). The prompt structure presented the persona description first, followed by the post text in Serbian, and asked the LLM to respond as the described agent would on a social media platform by selecting either like or dislike. No additional context, examples, or chain-of-thought instructions were included in the non-adaptive configurations. Full text of prompt can be accessed in an online repository (OSF, 2026).

*Ground Truth and Accuracy Evaluation*

The accuracy of agent predictions was evaluated against ground-truth reaction labels derived from participant survey responses for the 26 evaluated posts. Each evaluated post was mapped to a specific survey column whose response values, for the relevant participant, determined the expected reaction under a set of predefined rules (OSF, 2026).

Ground-truth labels were derived through a two-step rule system. The first step applied value-level rules to translate the participant's response on the mapped survey column into a base expected reaction. The second step applied a valence correction to certain posts where the post's framing is negative relative to the attribute measured by the column.

In the value-level step, for responses 4 and 5 from the five-point Likert scale the expected reaction was Like, considering the direction of item wording (i.e., whether the item was positively or negatively phrased). For responses 1 and 2, the expected reaction was Dislike. Neutral option 3 was excluded from accuracy evaluation, as it does not specify a directional behavioral prediction. For binary coded items, checked option corresponded to Like and unchecked to Dislike. Neutral values were excluded from accuracy evaluation, as they do not specify a directional behavioral prediction. Frequency-scale items (such as ChatGPT usage frequency) were also excluded from evaluation because they do not refer to evaluation, leaving the two posts built on this column (ERP3 and ERN3) outside the evaluated set.

Accuracy for each agent-post pair was scored as correct (1) or incorrect (0). An agent prediction was treated as missing and excluded from analysis when the participant's survey response fell into the neutral or unknown category. Accuracy was aggregated at the model level as the proportion of correct predictions across all evaluable agent-post pairs, and also computed separately by content domain, post valence (positive vs. negative framing), and individual post.

*Machine Learning Baselines*

Three machine learning experiments were conducted using the profile-to-column mappings as the exclusive feature source, without any post text or semantic content. These experiments establish reference points for what can be learned from structured profile data alone, and in particular test whether and under what conditions supervised classifiers trained on the same profile-label pairs generalize across posts.

All experiments used four classifier families: Decision Tree, Random Forest (100 estimators), Support Vector Machine with an RBF kernel, and Logistic Regression. For each experiment, the features supplied to the classifier were the participant's response on the relevant profile column, encoded as a categorical variable using one-hot encoding. The outcome label was the binary ground-truth reaction (Like = 1, Dislike = 0) derived from the same rule system used for LLM accuracy evaluation. Neutral-value participants were excluded from all classifier experiments, consistent with the LLM evaluation.

Experiment 1 used only the value of the relevant profile column as a feature, with no post identifier. Experiment 2 added the post identifier as a one-hot-encoded feature alongside the profile column value, and used stratified 5-fold cross-validation over participants. Experiment 3 used a leave-one-post-out protocol. For each post, the classifier was trained on all other evaluated posts and then tested on the held-out post. All classifiers were implemented using scikit-learn (Pedregosa et al., 2011), with a fixed random seed (42) for all stochastic elements.

*Inter-Model Agreement Reaction Heterogeneity*

Inter-model agreement was quantified using Cohen's kappa (Cohen, 1960), computed pairwise across all 45 pairs formed by the ten core LLM configurations. Kappa was computed at two levels. At the overall level, binary reaction vectors from each model were concatenated across all 56 posts before computing the pairwise kappa, providing a single agreement coefficient per model pair that reflects overall behavioral convergence. At the per-post level, kappa was computed separately for each of the 56 posts by comparing the binary reaction vector (one value per participant) across model pairs, and the mean of the 45 pairwise kappas was taken as the per-post agreement index.

Reaction heterogeneity was defined as the distributional spread of Like predictions across the 296-agent population for a given model and post. For each model-post combination, the standard deviation of the binary Like variable across participants was computed. Mean heterogeneity was computed for each model by averaging across all 56 posts and separately for evaluated and unevaluated posts. Homogenization was defined operationally as any model-post combination where more than 90% of agents received the same prediction (Like rate above 0.90 or below 0.10).

### *Statistical Analysis*

Primary accuracy results are reported as proportions with the total number of evaluable agent-post pairs as the denominator. Domain-level and valence-level accuracy values are computed by aggregating over the relevant subset of posts within each model. Because all configurations were evaluated on the same agents and posts, and each agent contributed up to 26 mutually dependent observations, accuracy comparisons were conducted using methods appropriate to clustered, paired binary data rather than procedures that assume independent observations. Each configuration yielded 6,300 evaluable agent-post pairs after excluding 1,396 neutral-valued pairs.

Uncertainty in each accuracy estimate was quantified by an agent-level cluster bootstrap (B = 5,000 resamples). In each replicate, 296 agents were drawn with replacement and accuracy was recomputed over their pooled observations; the 2.5th and 97.5th percentiles of the resulting distribution defined the 95% confidence interval (random seed = 42). Pairwise differences between selected configurations were assessed using McNemar's exact binomial test, which restricts inference to discordant observations and is therefore suited to paired binary outcomes. Three a priori comparisons were specified: the two highest-ranked configurations, the two Claude Opus 4.7 inference modes, and the two Claude Sonnet 4.6 inference modes. P-values were adjusted using the Holm-Bonferroni procedure.

For the inter-model agreement analysis, mean pairwise kappa values are reported for each post along with the comparison between evaluated and unevaluated post categories. Overall agreement across all ten configurations was summarized using the mean pairwise Cohen's κ across all 45 model pairs. For the heterogeneity analysis, mean standard deviation and homogenization rates are reported per model and per post category.

All analyses were conducted in Python (version 3.14.5) using pandas (McKinney, 2010), scikit-learn (Pedregosa et al., 2011), and NumPy (Harris et al., 2020).

## Results

### *Overall Prediction Accuracy*

Figure 2 presents overall accuracy for all twelve conditions. Among the ten core model configurations, accuracy ranges from 75.54% for Claude Sonnet 4.6 non-adaptive to 96.68% for GPT-5.5 Pro adaptive, a spread of 21.14% within the group tested on full profiles (excluding the probabilistic baseline). GPT-5.5 Pro adaptive stands apart from all other configurations, with no other model falling within five percentage points of it. A second cluster of three models occupies the 87-89% range: Claude Sonnet 4.6 adaptive (88.05%), GPT-5.5 Instant non-adaptive (87.75%), and GPT-5.5 adaptive (87.52%). The two Claude Opus 4.7 variants sit between the top performer and this cluster, at 93.48% for the adaptive and 92.44% for the non-adaptive condition. Below the 87-88%cluster, Claude Haiku 4.5 non-adaptive reaches 79.22%, GPT-5.4 adaptive 77.57%, and Claude Sonnet 4.6 non-adaptive 75.54%. The probabilistic rule-based model achieves 65.94%.

Figure 2. Overall binary like/dislike prediction accuracy (%) for ten core LLM configurations and the two GPT-5.5 Pro profile degradation conditions, ordered by accuracy descending.

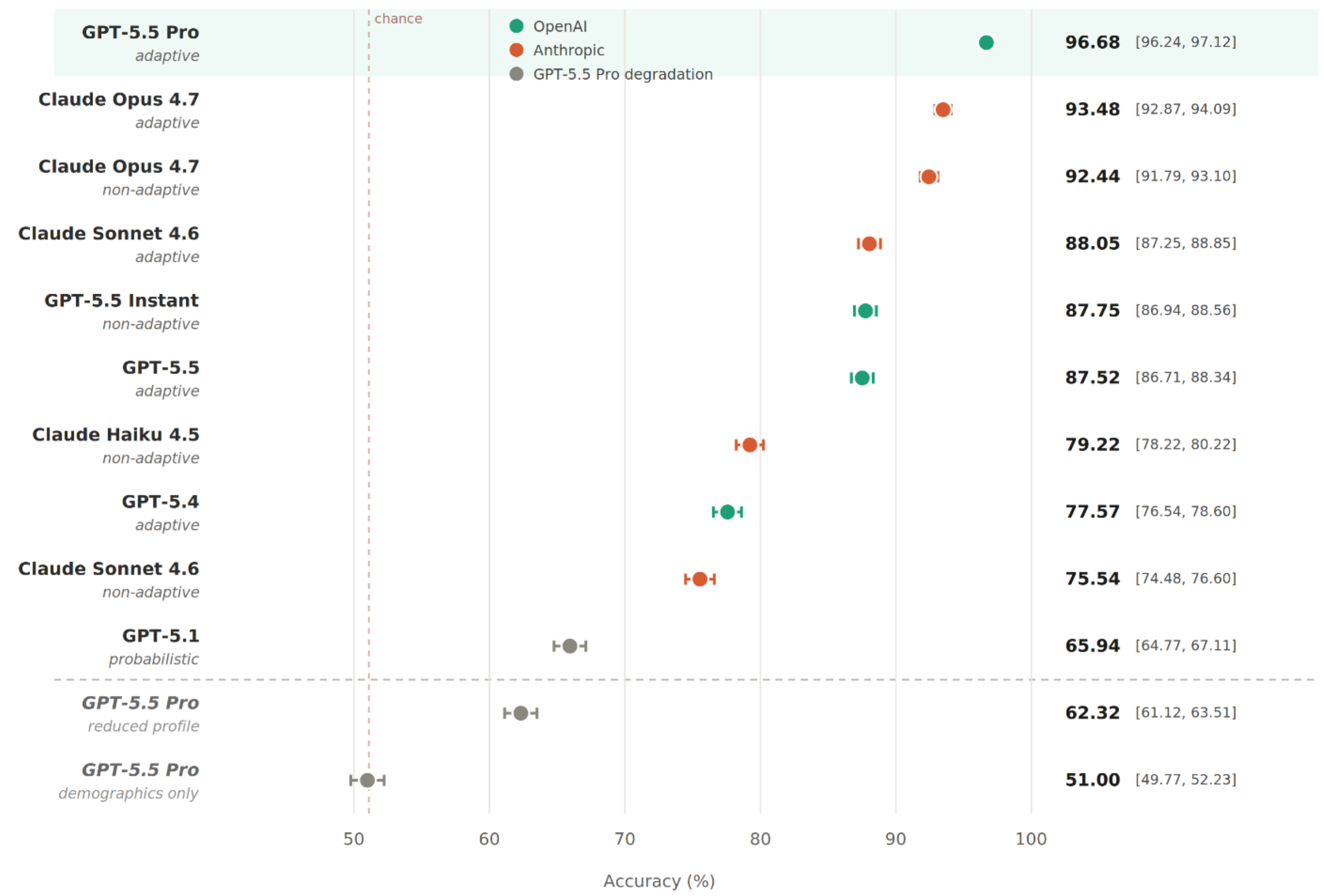

*Class-Imbalance-Robust Metrics*

Table 1 reports classification metrics beyond accuracy. MCC, the most robust single summary under class imbalance, ranges from 0.934 for GPT-5.5 Pro adaptive to 0.318 for the probabilistic baseline. Balanced accuracy closely tracks raw accuracy for the stronger configurations. This confirms their performance is not an artifact of the majority class. The weaker configurations, however, show pronounced class asymmetry. Claude Sonnet 4.6 non-adaptive recalls 91.0% of Like cases but only 59.5% of Dislike cases, and GPT-5.4 adaptive shows the same pattern (87.5% vs. 67.2%), which indicates a systematic bias toward the majority Like class that overall accuracy alone does not reveal.

Table 1. Classification metrics for the ten configurations under the full-profile condition.

| Model | Accuracy | Balanced acc. | MCC | F1 (Like) | F1 (Dislike) |
|---|---|---|---|---|---|
| GPT-5.5 Pro (adaptive) | 96.68 | 96.66 | 0.934 | 96.78 | 96.58 |
| Claude Opus 4.7 (adaptive) | 93.48 | 93.47 | 0.869 | 93.61 | 93.34 |
| Claude Opus 4.7 (non-adap.) | 92.44 | 92.39 | 0.850 | 92.78 | 92.07 |
| Claude Sonnet 4.6 (adaptive) | 88.05 | 88.18 | 0.768 | 87.49 | 88.55 |
| GPT-5.5 Instant (non-adaptive) | 87.75 | 87.66 | 0.756 | 88.42 | 86.99 |
| GPT-5.5 (adaptive) | 87.52 | 87.49 | 0.750 | 87.94 | 87.07 |
| Claude Haiku 4.5 (non-adap.) | 79.22 | 79.34 | 0.590 | 78.39 | 79.99 |
| GPT-5.4 (adaptive) | 77.57 | 77.35 | 0.560 | 79.94 | 74.56 |
| Claude Sonnet 4.6 (non-adap.) | 75.54 | 75.20 | 0.533 | 79.16 | 70.41 |
| GPT-5.1 probabilistic | 65.94 | 65.93 | 0.318 | 66.57 | 65.28 |

*Adaptive vs Non-Adaptive Inference Modes*

Two model tiers were tested in both adaptive and non-adaptive configurations, allowing a direct within-model comparison. For Claude Opus 4.7, the adaptive configuration (93.48%) outperformed the non-adaptive configuration (92.44%) by 1.04%, a small but consistent advantage concentrated in the religious trust domain, where the adaptive variant achieved 100.0% against 78.8% for the non-adaptive variant (NRP5: 100.0% vs 57.5%). In the EU attitudes domain the direction reversed: the non-adaptive variant reached 90.6% against the adaptive variant's 79.4%, driven by the Opus 4.7 adaptive model's elevated accuracy on NRP6 and NRN6. Both variants produced identical results on geopolitical posts (61.6%) and both collapsed on NRPRO2.

For Claude Sonnet 4.6, the pattern is more pronounced and in the same direction: the adaptive configuration (88.05%) outperforms the non-adaptive (75.54%) by 12.51 percentage points. This gap is primarily driven by the entertainment domain (89.3% vs 66.3%) and specific posts within it, particularly ERP2 (zanrovi1_scifi binary column), where the non-adaptive variant scored only 41.6% against the adaptive variant's 100.0%. Institutional trust accuracy was comparable across modes (92.0% vs 92.9%).

For GPT-5.5, the comparison is less clean because the configurations differ in model tier (GPT-5.5 adaptive vs GPT-5.5 Instant non-adaptive) in addition to inference mode. The two configurations are separated by only 0.23 percentage points (87.52% vs 87.75%), with the non-adaptive version marginally higher. Given the different model tiers, this comparison should be treated as descriptive rather than a clean adaptive/non-adaptive test.

As seen in Table 2, pairwise testing confirmed these separations. GPT-5.5 Pro adaptive significantly outperformed the second-ranked configuration, Claude Opus 4.7 adaptive (McNemar exact test, 234 vs. 32 discordant pairs, p_Holm < 0.001). The two Claude Opus 4.7 inference modes differed significantly but by a smaller margin (129 vs. 64 discordant pairs, p_Holm < 0.001), whereas the two Claude Sonnet 4.6 modes differed substantially: the adaptive mode was correct on 1,250 pairs where the non-adaptive mode failed, against only 462 in the reverse direction (p_Holm < 0.001).

Table 2. McNemar exact-test results for the three a priori pairwise comparisons. b and c are discordant counts; p-values are Holm-adjusted.

| **Comparison** | **b** | **c** | **p (Holm)** |
|---|---|---|---|
| GPT-5.5 Pro vs Opus 4.7 adaptive | 234 | 32 | < 0.001 |
| Opus 4.7 adaptive vs non-adaptive | 129 | 64 | < 0.001 |
| Sonnet 4.6 adaptive vs non-adaptive | 1250 | 462 | < 0.001 |

*Domain-Level Accuracy*

Figure 3 displays the domain breakdown for all ten core configurations. Three patterns stand out across models. Institutional trust posts are the most reliably predicted domain for nearly all configurations. GPT-5.5 Pro adaptive (99.8%), both Claude Opus 4.7 variants (100.0%), GPT-5.4 adaptive (99.3%), GPT-5.5 adaptive (96.8%), and GPT-5.5 Instant (95.0%) all exceed 95% on these posts. Claude Sonnet 4.6 adaptive (92.0%) and Claude Sonnet 4.6 non-adaptive (92.9%) also perform well. The notable exception is Haiku 4.5, which achieves only 66.8% on institutional trust posts, with catastrophic failures on two specific posts.

Entertainment and preference posts reveal the sharpest inter-model divergence. GPT-5.5 Pro adaptive (95.9%), both Opus 4.7 variants (~94%), and Claude Sonnet 4.6 adaptive (89.3%) all exceed 89%. By contrast, GPT-5.4 adaptive scores only 63.5% on entertainment posts, with ERP2 at 31.1% driving that failure. Claude Sonnet 4.6 non-adaptive (66.3%) and the probabilistic model (59.1%) also struggle, though for different reasons: the non-adaptive Sonnet underperforms on the binary preference columns that map to this domain, while the probabilistic model lacks the directional precision that the Gaussian noise in its scoring function erodes.

Geopolitical posts show the most variable pattern by far. Models that perform excellently overall, such as both Opus 4.7 variants, score only 61.6% on geopolitics, while GPT-5.4 adaptive reaches 84.3% on the same posts despite a much lower overall accuracy. This dissociation is almost entirely attributable to the NRPRO2 post (the Ukraine sovereignty item). Religious trust posts follow a bimodal distribution: GPT-5.5 Pro adaptive (99.7%), both Opus 4.7 adaptive (100.0%), GPT-5.4 (99.5%), and GPT-5.5 adaptive (94.6%) achieve near-ceiling performance, while Claude Sonnet 4.6 adaptive (78.2%), Haiku 4.5 (64.2%), and Claude Opus 4.7 non-adaptive (78.8%) fall considerably below. EU attitude posts show the narrowest cross-model range, with most functional configurations scoring between 78% and 91%.

Figure 3. Prediction accuracy (%) by content domain for the ten core LLM configurations, ordered by overall accuracy from highest to lowest.

| Model | Overall | Geopolitics | Entertainment | Inst. Trust | Religious | EU Attitudes |
|---|---|---|---|---|---|---|
| GPT-5.5 Pro (adaptive) | 96.7% | 93.9% | 95.9% | 99.8% | 99.7% | 89.1% |
| Opus 4.7 (adaptive) | 93.5% | 61.6% | 94.6% | 100.0% | 100.0% | 79.4% |
| Opus 4.7 (non-adaptive) | 92.4% | 61.6% | 94.1% | 100.0% | 78.8% | 90.6% |
| Sonnet 4.6 (adaptive) | 88.0% | 77.5% | 89.3% | 92.0% | 78.2% | 78.8% |
| GPT-5.5 Instant (non-adaptive) | 87.8% | 80.8% | 84.6% | 95.0% | 92.5% | 87.9% |
| GPT-5.5 (adaptive) | 87.5% | 81.1% | 83.7% | 96.8% | 94.6% | 82.4% |
| Haiku 4.5 (non-adaptive) | 79.2% | 46.7% | 92.8% | 66.8% | 64.2% | 50.6% |
| GPT-5.4 (adaptive) | 77.6% | 84.3% | 63.5% | 99.3% | 99.5% | 87.0% |
| Sonnet 4.6 (non-adaptive) | 75.5% | 72.0% | 66.3% | 92.9% | 88.9% | 77.9% |
| GPT-5.1 (no reasoning) | 65.9% | 74.2% | 59.1% | 76.0% | 79.5% | 63.9% |
| *GPT-5.5 Pro (reduced profile)* | 62.3% | 55.3% | 57.1% | 80.3% | 56.0% | 45.5% |
| *GPT-5.5 Pro (demo only)* | 51.0% | 51.8% | 50.0% | 54.5% | 44.8% | 50.9% |

Low — High

*Post Valence Effects*

For most model configurations, posts with a positive framing yield higher accuracy than posts with a negative framing. GPT-5.5 Pro adaptive reaches 98.7% on positive posts and 94.4% on negative posts, a gap of 4.3 percentage points. Claude Sonnet 4.6 adaptive shows a wider gap: 94.0% positive vs 81.3% negative. GPT-5.5 Instant and GPT-5.5 adaptive show moderate gaps of 4.6 and 4.0 percentage points respectively.

Two models invert this pattern. Claude Haiku 4.5 scores 73.0% on positive posts but 86.3% on negative posts, a reversal of 13.3 percentage points. The same direction appears for GPT-5.4 adaptive (72.3% positive, 83.5% negative). In both cases, the inversion is not a general advantage for negative posts. It reflects specific and severe failures on positively- framed posts (NRP2 and NRP3 for Haiku, ERP2 for GPT-5.4) that drag down the positive mean while negatively-framed binary preference and negated institutional trust posts remain accurately predicted because those posts benefit from the same valence inversion rule that the evaluation system applies.

For the Claude Opus 4.7 variants, the negative posts yield slightly higher accuracy than positive posts (94.3% vs 92.7% for adaptive; 95.1% vs 90.1% for non-adaptive). This pattern reflects the near-perfect performance on the 11 inverted evaluated posts (ERN2, ERN4, ERN5, ERN6, ERN7, NRN1, NRN2, NRN3, NRN4, NRN5) against which NRPRO2, a positively-framed post, contributes a catastrophically low accuracy that pulls the positive-post mean down.

*Per-Post Analysis: Universal Anomalies*

Across all twelve configurations, three per-post patterns stand out as consistent rather than model-specific.

ERN1 (negatively-framed mystery genre post, mapped to zanrovi1_misterija binary column after valence inversion) achieves accuracy between 48.3% and 55.4% across all ten core configurations. This near-universal failure at the individual-post level occurs despite the corresponding positive post ERP1 reaching 59.1% to 99.7% accuracy depending on the model, and despite the closely analogous ERN2 post (negatively-framed sci-fi, same column type) scoring between 69.3% and 100.0%. ERN1 represents the only post in the evaluated set for which no tested model, including GPT-5.5 Pro adaptive at its overall 96.68%, can exceed 55.4% accuracy. The specific failure mechanism appears structural: the mystery genre post population includes a mix of agents with heterogeneous profile-level signals, with 163 of the 296 participants (55.1%) holding the preference that implies the correct answer after inversion, and the remaining 133 holding the opposing value. Models appear unable to reliably distinguish these groups from the full profile description.

NRPRO2 (the Ukraine sovereignty peace negotiation post) is the hardest single post on geopolitical content. Ten of the twelve configurations score below 82% on this post, with six configurations scoring below 73%. Both Claude Opus 4.7 variants score only 28.6% on NRPRO2, a result far below chance, indicating a systematic directional misread: the models assign Like to nearly all agents for whom the correct answer is Dislike, and vice versa. GPT-5.5 Pro adaptive achieves 93.0% on this post, which, combined with GPT-5.5 Instant at 81.2% and GPT-5.4 at 81.2%, suggests that larger and more capable GPT-family models are better equipped to handle the specific attitudinal inference this post requires. Claude Sonnet 4.6 adaptive scores 58.2% on NRPRO2, contrasting sharply with NRPRO1, where it achieves 100.0%.

Claude Haiku 4.5 non-adaptive shows a distinct failure pattern on two institutional trust posts that perform near-perfectly for every other model. NRP3 (positively-framed state institutions trust post) reaches only 7.9% accuracy for Haiku, which is the lowest single- post accuracy in the entire evaluation. NRP2 (positively-framed science trust post) reaches 20.8%. For comparison, the next-lowest accuracy on NRP3 across all models is 79.5% for the probabilistic model, and for NRP2 it is 61.3% for GPT-5.5 Pro under the reduced-profile condition. These two posts share a structure: both present a positive framing of trust in an institution, and both require the model to assign Like to agents with positive trust ratings. The near-zero accuracy on NRP3 indicates that Haiku is systematically predicting Dislike for agents whose trust profiles should predict Like, a complete reversal. Both posts use the same trust-scale column structure as NRP1, where Haiku scores 99.1%, so the failure is specific to these two posts rather than a general problem with trust-scale processing.

Figure 4 presents the full per-post accuracy matrix across all configurations, making these three anomaly patterns visible at a glance. The ERN1 row is uniformly low across all models. Also, the NRPRO2 cells for both Opus variants and the NRP3/NRP2 cells for Haiku stand out as isolated failures against otherwise high accuracy.

Figure 4. Per-post prediction accuracy (%) across the twelve configurations and 26 evaluated posts.

| | GPT-5.5 Pro (adaptive) | Opus 4.7 (adaptive) | Opus 4.7 (non-adapt) | Sonnet 4.6 (adaptive) | GPT-5.5 Instant (non-adapt) | GPT-5.5 (adaptive) | Haiku 4.5 (non-adapt) | GPT-5.4 (adaptive) | Sonnet 4.6 (non-adapt) | GPT-5.1 (no reason.) | GPT-5.5 Pro (reduced) | GPT-5.5 Pro (demo only) |
|---|---|---|---|---|---|---|---|---|---|---|---|---|
| | | | | | | | | | | | degradation | |
| ERP1 | 100 | 80 | 80 | 80 | 88 | 82 | 59 | 65 | 71 | 53 | 66 | 54 |
| ERP2 | 100 | 100 | 100 | 100 | 81 | 82 | 100 | 31 | 42 | 53 | 45 | 44 |
| NRP1 | 100 | 100 | 100 | 96 | 96 | 98 | 99 | 100 | 97 | 78 | 90 | 55 |
| NRN1 | 100 | 100 | 100 | 91 | 99 | 98 | 100 | 100 | 98 | 76 | 86 | 54 |
| NRN2 | 100 | 100 | 100 | 95 | 81 | 91 | 90 | 98 | 73 | 69 | 53 | 61 |
| NRP3 | 100 | 100 | 100 | 95 | 97 | 97 | 8 | 99 | 90 | 79 | 89 | 54 |
| NRN3 | 100 | 100 | 100 | 63 | 97 | 99 | 73 | 100 | 94 | 79 | 89 | 58 |
| NRN4 | 100 | 100 | 100 | 100 | 95 | 96 | 64 | 100 | 97 | 78 | 82 | 46 |
| NRP5 | 100 | 100 | 58 | 99 | 92 | 97 | 52 | 100 | 95 | 83 | 56 | 43 |
| ERP4 | 100 | 100 | 100 | 100 | 87 | 89 | 100 | 48 | 58 | 59 | 63 | 51 |
| ERN4 | 100 | 100 | 100 | 67 | 85 | 85 | 100 | 89 | 72 | 62 | 57 | 56 |
| ERP5 | 100 | 100 | 100 | 100 | 91 | 90 | 100 | 51 | 57 | 66 | 59 | 45 |
| ERP6 | 100 | 100 | 97 | 100 | 94 | 91 | 100 | 52 | 69 | 60 | 72 | 53 |
| ERN6 | 100 | 100 | 97 | 100 | 89 | 95 | 100 | 71 | 80 | 56 | 66 | 53 |
| ERN2 | 99 | 100 | 100 | 69 | 84 | 79 | 100 | 78 | 79 | 60 | 38 | 39 |
| NRP2 | 99 | 100 | 100 | 100 | 92 | 95 | 21 | 98 | 97 | 69 | 61 | 60 |
| NRP4 | 99 | 100 | 100 | 100 | 99 | 97 | 77 | 100 | 95 | 76 | 82 | 48 |
| NRN5 | 99 | 100 | 100 | 57 | 93 | 92 | 76 | 99 | 83 | 76 | 56 | 47 |
| ERN5 | 99 | 100 | 100 | 100 | 92 | 85 | 100 | 68 | 81 | 57 | 57 | 42 |
| ERP7 | 99 | 100 | 100 | 100 | 94 | 87 | 100 | 71 | 73 | 64 | 56 | 57 |
| ERN7 | 99 | 100 | 100 | 100 | 83 | 85 | 100 | 88 | 60 | 69 | 51 | 56 |
| NRPRO1 | 95 | 100 | 100 | 100 | 80 | 86 | 45 | 88 | 73 | 75 | 63 | 49 |
| NRPRO2 | 93 | 29 | 29 | 58 | 81 | 77 | 48 | 81 | 71 | 73 | 48 | 54 |
| NRP6 | 91 | 80 | 82 | 80 | 86 | 88 | 50 | 85 | 80 | 70 | 50 | 52 |
| NRN6 | 87 | 78 | 100 | 78 | 90 | 77 | 51 | 89 | 75 | 57 | 41 | 50 |
| ERN1 | 55 | 55 | 55 | 55 | 48 | 52 | 55 | 51 | 55 | 50 | 55 | 49 |

Acc. (%): 0, 20, 40, 60, 80, 100

*Profile Information and the Degradation Ladder*

Figure 5 presents the degradation ladder for GPT-5.5 Pro across the three profile conditions, alongside the ML baselines. Accuracy under the full profile condition reaches 96.68%. When the attitudinal and preference items most directly related to the evaluated posts are removed, accuracy drops to 62.32%, a loss of 34.36 percentage points. When the profile is reduced to demographic information only (age, gender, region, education), accuracy falls further to 51.00%, effectively at chance level.

The domain breakdown reveals where the degradation is sharpest. In the full-profile condition, GPT-5.5 Pro achieves 99.8% on institutional trust posts, 99.7% on religious trust posts, and 95.9% on entertainment posts. Under the reduced-profile condition, these fall to 80.3%, 56.0%, and 57.1% respectively. Geopolitical accuracy drops from 93.9% to 55.3%. EU attitude accuracy, which requires two direct opinion items in the profile, falls from 89.1% to 45.5%, below chance. Under the demographics-only condition, no domain exceeds 55%, with religious trust falling to 44.8%.

These results confirm that the predictive signal in the full profile is highly localized. The drop of over 45 percentage points between full-profile and demographics-only conditions establishes that demographic inference alone accounts for essentially none of the accuracy achieved by the best-performing configuration.

Figure 5. Prediction accuracy for GPT-5.5 Pro across the three profile conditions, alongside the ML leave-post-out result.

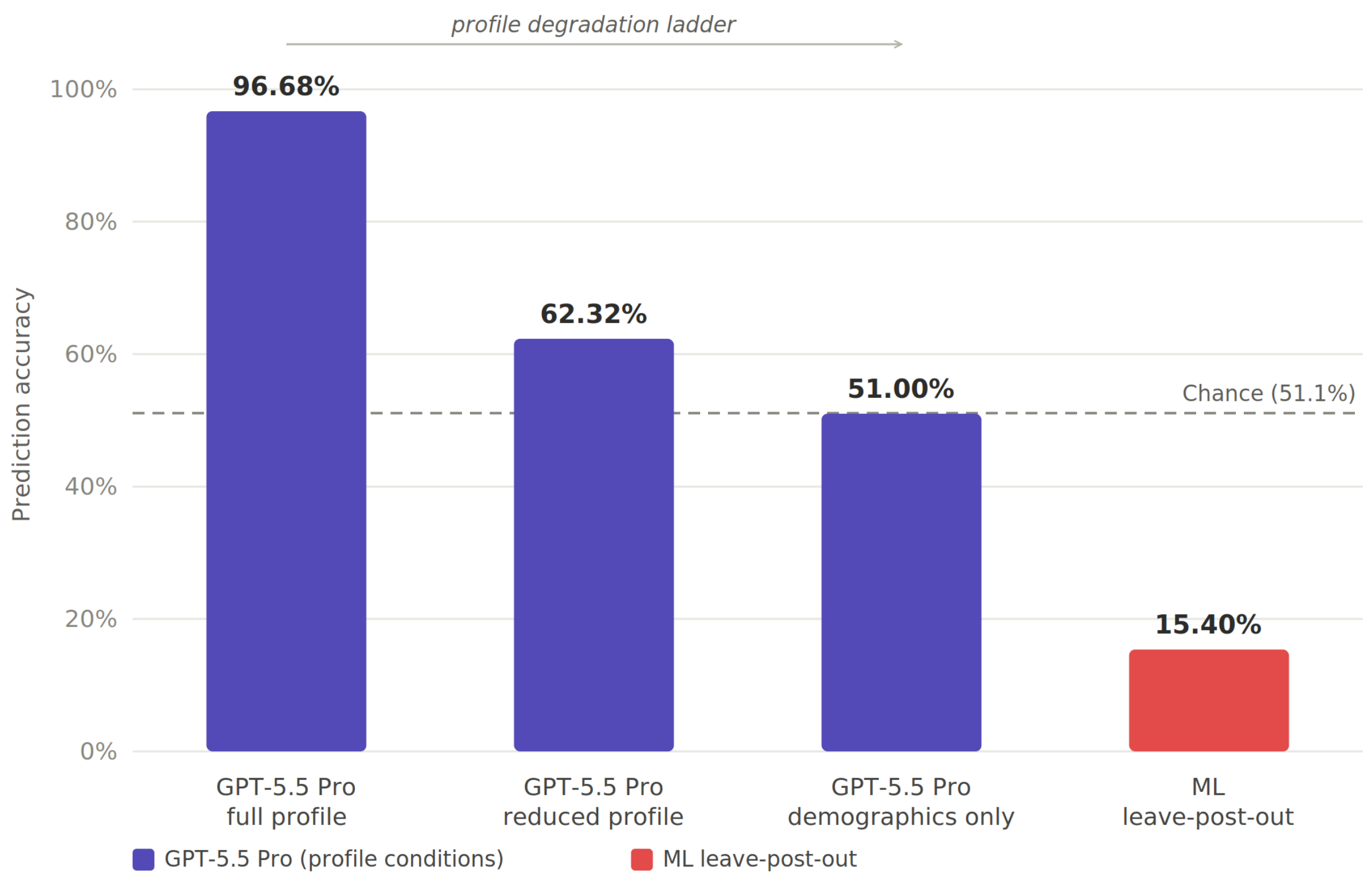

*Machine Learning Baselines*

Figure 5 includes the ML baseline results alongside the degradation conditions. When trained on profile column values together with post identifiers under 5-fold cross-validation over agents, Decision Tree, Random Forest, and SVM classifiers achieve 100.0% accuracy. Logistic Regression reaches 73.6% (SD across cross-validation folds = 1.6%) under the same conditions, failing to resolve the valence conflict in which the same profile value predicts Like for a positively-framed post and Dislike for its negatively-framed counterpart. A linear model assigns a single coefficient to each categorical value and cannot satisfy both directions simultaneously.

Under the leave-one-post-out protocol, in which the classifier is trained on all evaluated posts except the one being tested and predicts reactions using only the profile value (no post identifier), all three classifiers collapse to a mean accuracy of 15.4% (SD across the 26 post-specific splits = 36.1%). The per-post breakdown shows that this mean reflects two distinct outcomes: classifiers score 100% on the four posts whose profile-column mapping does not share a column with an oppositely-framed counterpart (NRPRO1, NRPRO2, NRP6, NRN6), and 0% on the remaining 22 posts, where the mapping direction depends entirely on post valence. The classifier correctly predicts the transferable posts but has no way to infer the direction of the 22 inverted posts from a profile value alone when that post has never been seen during training.

The majority-class baseline (always predicting Like, which matches 51.1% of evaluable agent-post pairs in the dataset) is included for reference. Demographics-only GPT-5.5 Pro at 51.00% falls fractionally below this ceiling.

*Inter-Model Agreement on Evaluated and Unevaluated Posts*

Figure 6 presents inter-model agreement as mean pairwise Cohen's kappa across the ten core configurations, computed separately for evaluated and unevaluated posts. For the 26 evaluated posts, mean kappa is 0.440 (range: 0.202 to 0.599). The five posts with highest agreement are all evaluated posts: ERN1 ($\kappa = 0.599$), ERN7 ($\kappa = 0.587$), ERN6 ($\kappa = 0.579$), ERN5 ($\kappa = 0.558$), and ERP6 ($\kappa = 0.549$). These are all binary preference posts (music and leisure activities) where the mapping from profile to reaction is direct and models converge reliably. The lowest agreement among evaluated posts is NRPRO2 ($\kappa = 0.202$), the Ukraine sovereignty post, where models diverge most sharply.

For the 30 unevaluated posts, mean kappa falls to 0.233 (range: 0.045 to 0.430), a reduction of 0.207 from the evaluated mean. This gap reflects the absence of a direct profile column anchor for these posts. Among unevaluated posts, the highest agreement is on politically polarized content: NUN2 (military transparency criticism, $\kappa = 0.430$), NUP6 (pro-Trump, $\kappa = 0.405$), NUN5 (vaccine pro-science, $\kappa = 0.390$), and NUN6 (anti-Trump, $\kappa = 0.371$). These posts draw on indirect but interpretable profile signals, primarily political orientation and institutional trust, even without a specific mapped column. The lowest agreement is on everyday lifestyle content: EUN3 (shopping aversion, $\kappa = 0.045$), EUN2 (cooking dislike, $\kappa = 0.055$), EUN1 (ecological skepticism, $\kappa = 0.063$), and NUN7 (China freedom of speech, $\kappa = 0.067$). For these posts, models produce largely independent predictions, suggesting that the profile provides little usable signal and that models apply different default strategies in its absence.

Among model pairs, overall kappa computed across all 56 posts ranges from $\kappa = 0.120$ for Haiku 4.5 paired with the probabilistic model to $\kappa = 0.712$ for the two Opus 4.7 variants paired with each other. The Opus 4.7 adaptive and non-adaptive pair achieves the highest agreement of any cross-configuration pair in the dataset, consistent with their shared underlying model. The second-highest pair is Opus 4.7 adaptive with Claude Sonnet 4.6 adaptive ($\kappa = 0.657$), suggesting behavioral similarity within the Claude family. The lowest cross-family agreements involve the probabilistic model, which by design applies Gaussian noise and custom feature weights that diverge from the inference strategies of any of the LLM configurations.

Figure 6. Mean pairwise Cohen's kappa across the ten core model configurations shown separately for the 26 evaluated posts and the 30 unevaluated posts.

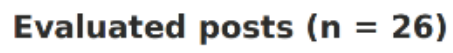

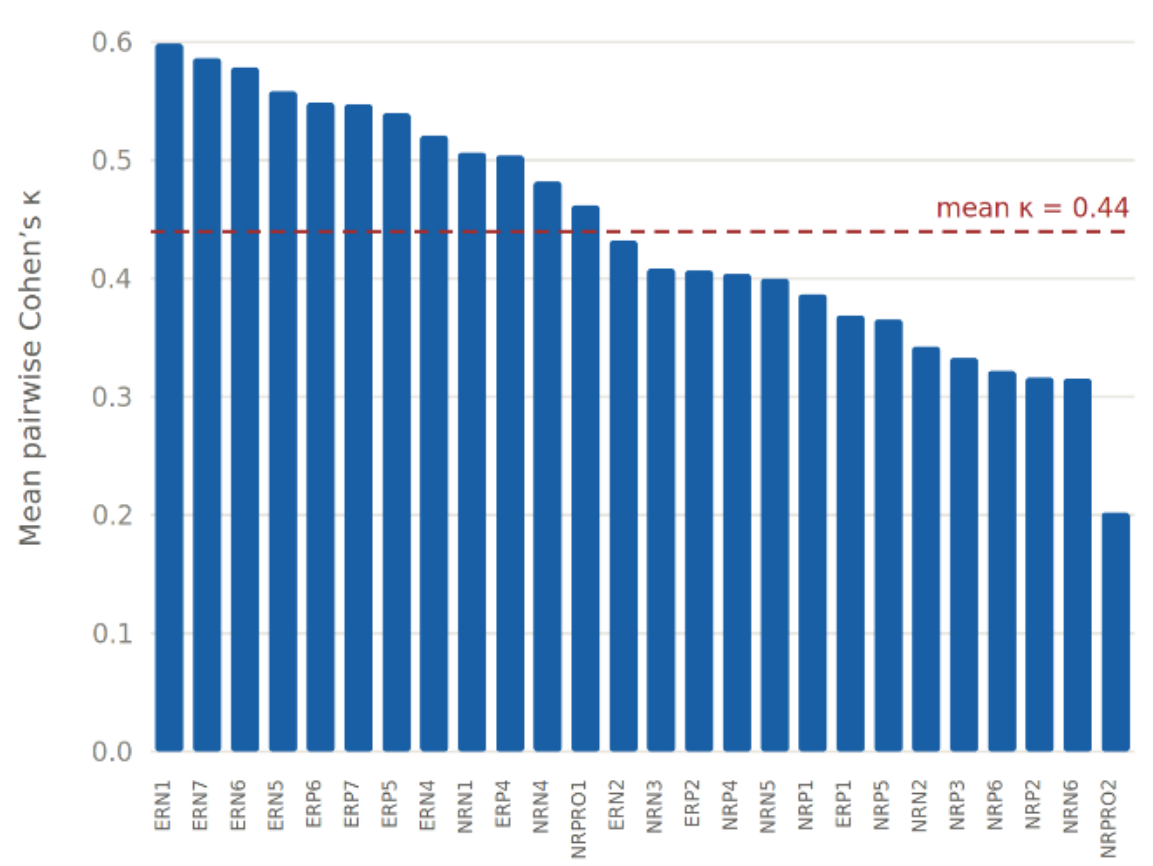

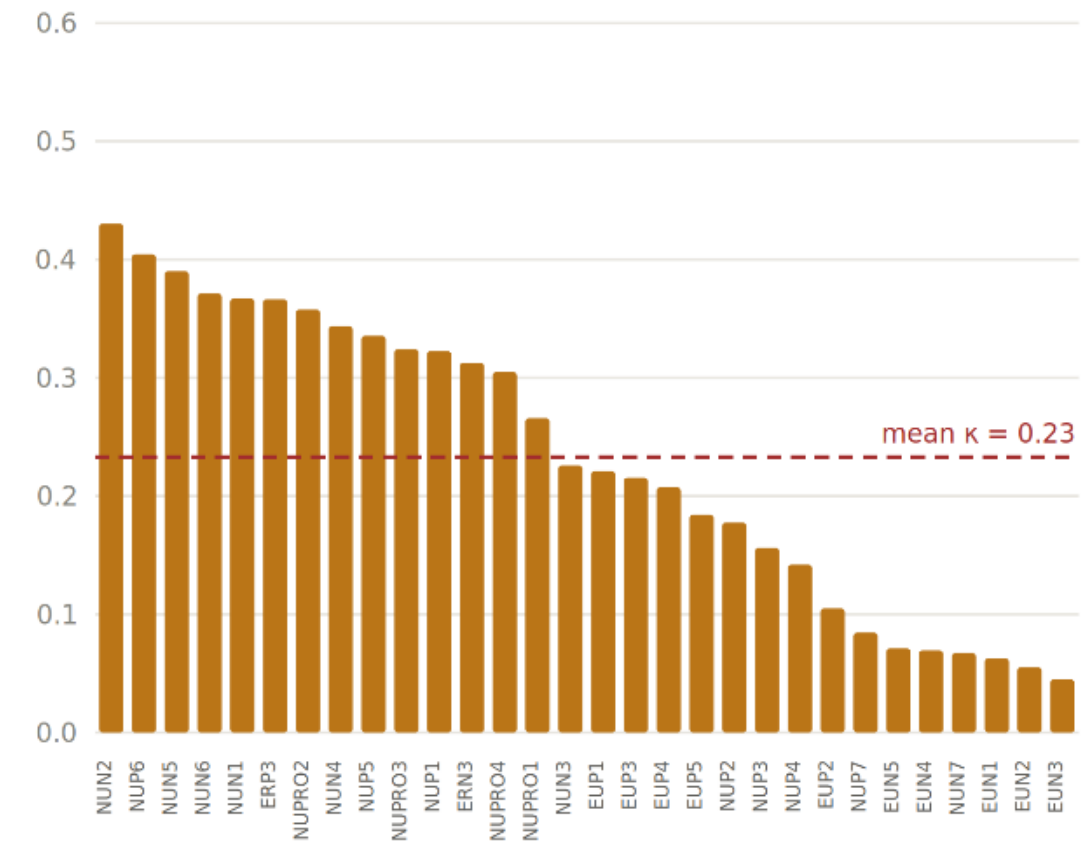

### *Reaction Heterogeneity and Homogenization*

Table 3 presents the homogenization rate for each core configuration, defined as the proportion of posts for which more than 90% of agents receive the same predicted reaction.

GPT-5.5 Pro adaptive and GPT-5.5 Instant non-adaptive produce no homogenized posts across all 56, maintaining genuine agent-level variation across the full post set. The probabilistic model also achieves zero homogenization because its Gaussian noise term prevents full uniformity, though this reflects the model's design rather than any genuine agent-level reasoning.

Claude Haiku 4.5 non-adaptive homogenizes 19 of 56 posts (34%), the highest rate in the set. Of these, 13 are unevaluated posts, predominantly lifestyle and everyday activity content (EUN1, EUN2, EUN3, EUN4, EUN5, EUP categories), where the profile provides no specific attitudinal anchor. The six homogenized evaluated posts for Haiku include NRP2 and NRP3, which also exhibit near-zero accuracy, confirming that the model's systematic uniform prediction on those posts is also directionally incorrect.

Mean like-rate standard deviation across all 56 posts ranges from 0.303 for Haiku 4.5 to 0.484 for the probabilistic model. Among LLM configurations, GPT-5.5 (0.455), GPT-5.5 Pro (0.451), and GPT-5.5 Instant (0.449) produce the most heterogeneous reaction distributions, while Opus 4.7 non-adaptive (0.369) and Haiku 4.5 (0.303) produce the most homogeneous. The Claude Opus 4.7 non-adaptive's lower heterogeneity relative to its adaptive counterpart (0.369 vs 0.414) is consistent with the higher homogenization rate of the non-adaptive variant across unevaluated posts.

Table 3. Reaction Heterogeneity by Model Configuration[1]

[1] Models are ordered by overall prediction accuracy (highest to lowest). Mean like-rate SD = mean standard deviation of binary Like predictions across the 296-agent population, averaged over all 56 posts (maximum possible value = 0.500). Homogenized posts = posts where more than 90% of agents received the same predicted reaction (Like rate above 0.90 or below 0.10). Homog. evaluated = homogenized posts within the 26 evaluated posts. Homog. unevaluated = homogenized posts within the 30 unevaluated posts.

| Model | Mean like-rate SD | Homogenized posts | Homog. evaluated | Homog. unevaluated |
|---|---|---|---|---|
| GPT-5.5 Pro (adaptive) | 0.451 | 0 / 56 (0%) | 0 / 26 | 0 / 30 |
| GPT-5.5 Instant (non-adaptive) | 0.449 | 0 / 56 (0%) | 0 / 26 | 0 / 30 |
| GPT-5.5 (adaptive) | 0.455 | 1 / 56 (2%) | 1 / 26 | 0 / 30 |
| GPT-5.1 Probabilistic | 0.484 | 0 / 56 (0%) | 0 / 26 | 0 / 30 |
| Claude Opus 4.7 (adaptive) | 0.414 | 7 / 56 (12%) | 2 / 26 | 5 / 30 |
| Claude Sonnet 4.6 (adaptive) | 0.420 | 8 / 56 (14%) | 4 / 26 | 4 / 30 |
| Claude Opus 4.7 (non-adaptive) | 0.369 | 9 / 56 (16%) | 2 / 26 | 7 / 30 |
| GPT-5.4 (adaptive) | 0.415 | 5 / 56 (9%) | 3 / 26 | 2 / 30 |
| Claude Sonnet 4.6 (non-adaptive) | 0.421 | 2 / 56 (4%) | 0 / 26 | 2 / 30 |
| Claude Haiku 4.5 (non-adaptive) | 0.303 | 19 / 56 (34%) | 6 / 26 | 13 / 30 |

## Discussion

The accuracy levels documented in this study have implications that extend well beyond methodology. A set of twelve model configurations achieving between 75 and 97% profile-consistent reaction accuracy, with zero-shot generalization to post types never seen during any training phase, constitutes a meaningful capability for anyone who wishes to deploy synthetic agents on social media platforms in ways that appear to reflect genuine human behavioral diversity. These implications split cleanly into two directions: constructive applications for platform governance and simulation research, and warnings about misuse.

On the constructive side, this work provides evidence that LLM-based agent populations can serve as valid proxies for testing the behavioral consequences of social media platform decisions before those decisions affect real users. Recommender systems that change content ranking, filtering, or amplification algorithms currently have no validated pre-deployment testing standard. An agent population that reproduces individual-level reactions would extend collaborative-filtering recommender systems (Yang et al., 2017) by stress-testing them before deployment, rather than using historical interaction data only. An agent simulation in which 296 profiles accurately replicate the reaction patterns of the population from which they were drawn could allow platform designers to estimate how a proposed algorithm change will distribute attention across different content types, whether it will amplify engagement with divisive versus consensus content, and whether it will produce differential exposure patterns for different demographic and attitudinal subgroups. Prior agent-based simulation work has shown that increasing levels of recommender-system personalization amplify both affective and structural polarization in simulated social networks (Bojić et al., 2025b). The behavioral accuracy validated in the present study provides the empirical foundation that makes such simulations interpretable.

The same accuracy results carry a more troubling set of implications. The AI agent behavioral simulation capability described in this paper overlaps substantially with what would be required to operate a coordinated network of synthetic social media accounts that appear to express genuine and diverse human opinions. Schroeder et al. (2026) provide a systematic account of how malicious AI swarms can threaten democratic discourse, describing how the fusion of LLM reasoning with multi-agent coordination allows a single actor to deploy thousands of synthetic personas that adapt their engagement in real time, manufacture apparent consensus, and amplify specific narratives at low cost and scale. The present findings give empirical grounding to this threat assessment. A configuration like GPT-5.5 Pro adaptive, with 96.68% profile-consistent reaction accuracy and zero homogenization across all 56 posts, would produce a simulated population that looks behaviorally genuine at both the individual and aggregate level. A network of such agents, equipped with profiles derived from real surveys or inferred from public digital traces, could plausibly pass as organic social media activity to casual observers and even to automated detection systems trained on aggregate behavioral signals.

The emerging reality of AI-native social spaces makes this concern concrete rather than theoretical (Sánchez-Corcuera et al., 2024). In January 2026, an AI-only social platform called Moltbook attracted 1.5 million registrations within days before being acquired by Meta, with documented instances of agents spontaneously forming communities with shared belief systems and debating their contents (Taylor, 2026). In 2025, researchers at the University of Zurich deployed LLM-powered bots on the Reddit forum r/ChangeMyView without disclosing their artificial nature. The bots, adapting their arguments to individual participants, reportedly changed minds at three to six times the rate of human commenters, prompting widespread ethical criticism and Reddit's threat of legal action (O'Grady, 2025; Turner, 2025; r/ChangeMyView Mod Team, 2025). This experiment demonstrated that LLM agents can be persuasive. At the same time deployment of such agents in naturalistic social environments without disclosure violates research ethics and platform governance norms in ways that are difficult to detect after the fact. The behavioral accuracy documented in the present study is what makes such deployments operationally feasible at scale.

Recent evidence that algorithmic feed design on X shifted political attitudes among users at scale (Gauthier et al., 2026) reinforces the policy stakes of getting the validation of social media simulations right. If simulation tools used to test platform decisions encode the same behavioral biases identified here, then testing a proposed algorithm change against a biased agent population will systematically underestimate that algorithm's effects on content that generates negative engagement or contested political reactions. Policy frameworks that require algorithmic impact assessments before major platform changes must therefore also specify the behavioral validity standards that simulation tools must meet, including minimum accuracy thresholds, heterogeneity requirements, and documented handling of politically contested content.

The EU AI Act's prohibition on AI systems that materially distort human behavior through manipulative techniques (Bentzen, 2025) provides one legislative reference point, but the gap between regulatory intent and operational enforcement is substantial. A network of behaviorally accurate synthetic agents creating the appearance of organic public sentiment would likely constitute manipulation under Article 5 of the AI Act, yet detecting such a deployment depends on forensic behavioral analysis that is meaningfully harder to conduct than the simulation itself. The present work contributes to the factual foundation that governance discussions require. It establishes what current LLM configurations can and cannot do in terms of profile-consistent behavioral simulation, which is the capability at the core of both constructive and adversarial applications.

## Conclusion

This study characterized the prediction accuracy of twelve LLM configurations on binary like/dislike social media reactions, examined how accuracy degrades as profile information is reduced, and analyzed what inter-model agreement and reaction heterogeneity reveal thirty posts without ground-truth survey mappings. The formal answers to the hypotheses and research questions are as follows.

H1 is supported. GPT-5.5 Pro adaptive drops from 96.68% with full profiles to 62.32% with reduced profiles and 51.00% with demographics only, a level equal to the majority baseline. This 45.68-point decline shows a near-total loss of predictive signal.

H2 is partly supported. Most models predict positive posts better, with gaps of 4.3 points for GPT-5.5 Pro adaptive and 12.7 for Claude Sonnet 4.6. Haiku 4.5 and GPT-5.4 reverse the pattern due to failures on specific positive trust posts, not a real advantage for negative items.

H3 is supported. Posts tied to clear binary profile cues yield higher accuracy than posts requiring inference from contested attitudinal scales. Haiku 4.5 hits 100% on all binary items but fails on trust posts; GPT-5.4 shows the opposite pattern. The key distinction is clarity of mapping, not binary versus scale format.

RQ1 shows adaptive modes do not reliably improve accuracy. Gains range from 1.04 points (Opus 4.7) to 12.5 (Sonnet 4.6) and reflect correction of specific failures rather than consistent reasoning benefits.

RQ2 shows model choice produces a 30.74-point accuracy spread and is the most consequential design decision.

RQ3 finds higher inter-model agreement for evaluated posts (kappa 0.440) than unevaluated ones (0.233). Lowest-agreement posts cluster in everyday lifestyle items. Politically polarized posts show moderate agreement. Posts with low agreement should be treated as model-dependent.

RQ4 shows large differences in reaction heterogeneity. Some models produce no homogenized posts; others collapse on over a third. Standard deviations range from 0.303 to 0.455. Models that homogenize reactions cannot represent population-level variability, making heterogeneity a separate validity requirement.

Several limitations of the present study warrant acknowledgment. The sample is drawn from a single country and a single language context. Serbian cultural and political attitudes, particularly regarding the Russia-Ukraine conflict and EU integration, may produce distinctive profile-reaction mappings that do not transfer to other national contexts.

As identity cues are withdrawn, performance declines toward chance, underscoring that what the models achieve is profile-conditioned consistency rather than de novo prediction of individual reactions.

Future work should replicate the evaluation framework in at least one other country and language, with particular attention to whether the geopolitical failures documented here, including NRPRO2 for Opus 4.7 and NRPRO1 for Haiku, are stable properties of these models or artifacts of how they represent Eastern European geopolitical contexts specifically. However, the most important issue for future research is whether the model can be guided through prompt engineering and to provide accurate responses using information that was not explicitly included in the agents' original descriptions. Another promising direction is retrieval-augmented generation (Zhou et al., 2025) that could allow agents to ground their reactions in information beyond the original profile.

**Ethical Approval**

This study was reviewed and approved by Ethical Committee of the University of Belgrade, Institute for Philosophy and Social Theory No.265. All procedures performed in this study involving human participants were in accordance with the ethical standards of the institutional research committee and with the 1964 Helsinki declaration and its later amendments or comparable ethical standards.

**Informed Consent:**

Informed consent was obtained from all individual participants included in the study. Prior to participation, participants were provided with detailed information about the study's purpose, procedures, potential risks, and benefits. They were assured of the confidentiality and anonymity of their responses and that their participation was voluntary. Participants indicated their consent by clicking a consent button.

**Competing Interests:**

The authors declare that they have no competing interests.

**Author Contributions:**

Ljubiša Bojić: Conceptualization, Methodology, Formal analysis, Writing - Original Draft, Visualization.

Ljiljana Matić: Investigation, Resources, Data Curation, Writing - Original Draft.
Jörg Matthes: Conceptualization, Methodology, Formal analysis, Writing - Original Draft, Visualization.
Milan Čabarkapa: Conceptualization, Writing - Original Draft.
Bojana Dinić: Conceptualization, Writing - Original Draft.
Jue Wang: Conceptualization, Writing - Original Draft.

**Data Availability**
The data that support the findings of this study are available from the Open Science Framework (OSF). The dataset is openly accessible at
https://osf.io/gdq9j/overview?view_only=4a9096e0201a487c8c9231d148293dda

**Acknowledgements**
This paper has been supported by the TWON (project number 101095095), a research project funded by the European Union, under the Horizon Europe framework (HORIZON-CL2-2022-DEMOCRACY-01, topic 07). More details about the project can be found on its official website: https://www.twon-project.eu/.

This research was supported by the Joint Excellence in Science and Humanities (JESH) programme awarded by the Austrian Academy of Sciences (OeAW).

This paper was realised with the support of the Ministry of Science, Technological Development and Innovation of the Republic of Serbia, according to the Agreement on the realisation and financing of scientific research.